\documentclass[apj]{emulateapj}

\usepackage{graphicx,times}
\usepackage{subfigure}
\newcommand{\be}{\begin{equation}}
\usepackage{threeparttable}
\usepackage{booktabs}
\newcommand{\ee}{\end{equation}}
\newcommand{\bea}{\begin{eqnarray}}
\newcommand{\eea}{\end{eqnarray}}

\usepackage{enumerate}
\usepackage{amsmath}
\usepackage{cases}
\usepackage{longtable}
\usepackage{hyperref}
\usepackage{epstopdf}
\usepackage{amsmath,bm}
\usepackage{amssymb}
\usepackage{natbib}
\usepackage{morefloats}
\usepackage{multirow}
\usepackage{array}
\usepackage{verbatim}

\begin{document}

\title{Magnetic field amplification in supernova remnants}

\author{Siyao Xu\altaffilmark{1,2} and Alex Lazarian\altaffilmark{1} }

\altaffiltext{1}{Department of Astronomy, University of Wisconsin, 475 North Charter Street, Madison, WI 53706, USA; 
sxu93@wisc.edu,
lazarian@astro.wisc.edu}
\altaffiltext{2}{Hubble Fellow}

\begin{abstract}

Based on the new findings on the turbulent dynamo in \citet{XL16}, 
we examine the magnetic field amplification in the context of supernova remnants. 
Due to the strong ion-neutral collisional damping in the weakly ionized interstellar medium, 
the dynamo in the preshock turbulence remains in the damping kinematic regime, which leads to 
{\it a linear-in-time growth of the magnetic field strength}. 
The resultant magnetic field structure 
enables effective diffusion upstream and shock acceleration of cosmic rays to energies above the ``knee". 
Differently,
the nonlinear dynamo in the postshock turbulence leads to
{\it a linear-in-time growth of the magnetic energy} due to the turbulent magnetic diffusion. 
Given a weak initial field strength in the postshock region, 
the magnetic field saturates 
at a significant distance from the shock front as a result of the inefficiency of the nonlinear dynamo. 
This result is in a good agreement with existing numerical simulations and 
well explains the X-ray spots detected far behind the shock front.

\end{abstract}


\section{Introduction}

Supernova remnants (SNRs) are the most plausible sources of Galactic cosmic rays (CRs)
\citep{Bla87}.
Magnetic-field amplification is expected in SNRs to ensure an efficient diffusive shock acceleration (DSA) 
\citep{Ax77, Bel78},
and is also supported by observational evidence, e.g., the magnetic fields on the order of $100~ \mu$G
near the shock front inferred from narrow X-ray synchrotron rims
\citep{Bam03, Bam05, Bamb05, Vin12},
the milligauss magnetic fields suggested by the rapid X-ray variability from compact sources in the downstream region 
\citep{Pat07, Uch07, Uch08}. 
The milligauss fields are also indicated from radio observations
\citep{Lon94}.
Clearly, the amplified magnetic fields in SNRs cannot be accounted for by the shock compression of the interstellar field strength of a few microgauss.

There are extensive studies 
on the origin of the magnetic fluctuations for confining CRs near the shock, e.g., 
the non-resonant streaming instability for driving magnetic fields at length scales smaller than the CR Larmor radius
\citep{Bell2004},
an inverse cascade of Alfv\'{e}n waves excited at the Larmor radius to larger wavelengths
\citep{Diamond_Makov}.
More generally, turbulence 
is believed to be an efficient agent to amplify magnetic fields via the turbulent dynamo 
\citep{Kaza68,KulA92}, 
which has also been invoked for explaining both the preshock 
(\citealt*{BJL09}, hereafter BJL09; \citealt{Dru12, Del16})
and postshock 
\citep{Bal01,Giac_Jok2007, Ino09,Guo12,Fra13,Ji16}
magnetic fields. 
As turbulence is induced by the interaction between SNR shocks and interstellar turbulent density fluctuations, 
the turbulent dynamo is inevitable in SNRs with the turbulent kinetic energy dominating over 
the pre-existing magnetic energy.

The theoretical advances in magnetohydrodynamic (MHD) turbulence have been made since 
\citet{GS95} 
and later works 
\citep{LV99, CV00, MG01}
with the conceptual improvement by introducing the {\it local} system of reference.
They bring new physical insights into the turbulent dynamo problem. 
Within the framework of the 
\citet{GS95}
model of MHD turbulence, 
a detailed analytical study on the turbulent dynamo process in plasmas with arbitrary conducing and ionization degrees 
was carried out by 
\citet{XL16} (hereafter XL16). 
A remarkable finding there is that the kinematic dynamo in a weakly ionized medium leads to a linearly growing field strength with time
\footnote{This should not be confused with the linearly growing magnetic energy with time that characterizes the 
nonlinear turbulent dynamo regime (see the following text).},  
and the resulting characteristic scale of the magnetic field can significantly exceed the viscous scale of turbulence.
This new dynamo regime is  referred to as the ``damping regime".
Since the interstellar media (ISM) that SNR shocks sweep through are frequently partially ionized 
\citep{Drai11},
the significant modifications on the kinematic dynamo in the presence of neutrals should be incorporated 
when studying the magnetic field amplification in the preshock region and its implications on CR acceleration.

For CR acceleration at shocks, 
strong magnetic fields in both the pre- and post-shock regions are necessary to trap and mirror CR particles to facilitate many 
shock crossings. 
The amplification of the preshock magnetic field is crucial.
It has been modelled in earlier studies 
(e.g. BJL09; \citealt{Dru12, Del16}) in an ideal situation with a fully ionized upstream plasma, 
and the amplification time is rather limited given a relatively thin precursor
(BJL09).
To reach a more realistic and generalized description, 
here we consider the partial ionization of the ISM and examine its influence on the CR diffusion 
in the amplified magnetic field in the CR precursor.

In the highly ionized postshock medium, the turbulent energy can cascade down to quite small scales. 
Very likely, the turbulent dynamo starts with an equipartition between the magnetic and turbulent energies at 
an intermediate scale and fall in the nonlinear regime.
Different from  
the kinematic dynamo with a strong dependence on the microscopic magnetic diffusion,
the nonlinear dynamo, which is initiated by the equipartition between the magnetic energy and the local turbulent energy, 
is mainly subject to the turbulent magnetic diffusion,
and consequently the magnetic energy has a linear-in-time growth with a low dynamo efficiency. 
This theoretical result in XL16 quantitatively agrees with earlier numerical studies, e.g., 
\citet{CVB09},
\citet{Bere11}.

Motivated by the enhanced field strength in SNRs indicated by observations, 
in this work, 
we investigate the magnetic field amplification in both the pre- and post-shock regions of a SNR 
by applying the general turbulent dynamo theory developed by XL16 and discuss its implications on the CR acceleration. 
In Section 2, we analyze the magnetic field amplification in the weakly ionized preshock medium and its implication on the 
CR diffusion upstream. 
In Section 3, we study the magnetic field amplification and CR diffusion in the fully ionized postshock medium. 
Discussions on an alternative acceleration mechanism of CRs in the 
postshock MHD turbulence are in Section 4. 
In Section 5, we summarize the main results.

\section{Magnetic field amplification in the partially ionized preshock region}

The turbulence in the preshock region can result from 
the CR pressure gradient in the shock precursor interacting with the density inhomogeneities in the upstream ISM
(BJL09).
The pre-existing interstellar field of a few $\mu$ G is expected to be amplified by the resulting turbulence 
via the small-scale turbulent dynamo process. 
The dynamo growth of magnetic field is driven by the turbulent motions. 
They are essentially hydrodynamic over the length scales larger than the equipartition scale, where the turbulent and magnetic energies are in equipartition, 
and are assumed to follow the Kolmogorov scaling.

In the partially ionized ISM, the dynamo evolution of magnetic field in the linear regime has its time-dependence and growth rate 
strongly affected by the ion-neutral collisional damping 
(\citealt{KulA92}; XL16).
For the propagation of a strong shock wave through the partially ionized ISM, it is necessary to take into account 
the partial ionization of the upstream medium for a realistic description of the magnetic field amplification in the preshock region.

\subsection{Damping effect on the kinematic dynamo}

The damping rate of magnetic fluctuations due to ion-neutral collisions is given by 
\citep{Kulsrud_Pearce, KulA92},
\begin{equation}\label{eq: damrat}
     \omega_d =   \mathcal{C} k^2 \mathcal{E}_M,   ~~ \mathcal{C} = \frac{\xi_n}{3\nu_{ni}},
\end{equation}
where $\mathcal{E}_M$ is the magnetic-fluctuation energy per unit mass, $\xi_n = \rho_n/\rho$
is the neutral fraction with the neutral mass density $\rho_n$ and the total mass density $\rho$, 
and $\nu_{ni}=\gamma_d \rho_i$ is the neutral-ion collision frequency with $\gamma_d$ as the drag coefficient  
(see \citealt{Shu92})
and $\rho_i$ as the ion mass density.

Magnetic fluctuations in strongly coupled neutrals and ions are also subject to the neutral-viscous damping
\citep{LVC04,XLY14}.
The corresponding viscous scale is
\begin{equation}\label{eq: knu}
 k_\nu=L^{-\frac{1}{4}}V_L^{\frac{3}{4}} \nu_n^{-\frac{3}{4}},
\end{equation}
where the Kolmogorov scaling of turbulence is used, $V_L$ is the turbulent velocity at the injection scale $L$,
$\nu_n=v_{th}/(n_n\sigma_{nn})$ is the kinematic viscosity in neutrals, with the neutral number density
$n_n$, the neutral thermal speed $v_{th}$, and the cross section of a neutral-neutral collision $\sigma_{nn}$.
If the ion-neutral collisional damping rate at $k_\nu$, 
$ \omega_d (k_\nu) =   \mathcal{C} k_\nu^2 \mathcal{E}_M$, 
is larger than the viscous damping rate, 
\begin{equation}
    \omega_v (k_\nu)= k_\nu^2 \nu_n = L^{-\frac{1}{2}} V_L^\frac{3}{2} \nu_n^{-\frac{1}{2}},
\end{equation}
that is 
\begin{equation} \label{eq: pram1}
    \mathcal{E}_M > \mathcal{C}^{-1} \nu_n, 
\end{equation}
ion-neutral collisions dominate over the neutral viscosity, leading to a damping scale of magnetic fluctuations 
larger than the viscous scale.

The turbulent eddies at the damping scale $k_d$ are mainly responsible for driving the dynamo growth of magnetic energy, 
at a rate comparable to the eddy-turnover rate, 
\begin{equation}\label{eq: gamlak}
   \Gamma_d= k_d v_d = L^{-\frac{1}{3}}V_Lk_d^\frac{2}{3},
\end{equation}
where the turbulent velocity at $k_d$ is, 
\begin{equation}\label{eq: vkd}
   v_d = V_L (k_d L)^{-\frac{1}{3}}. 
\end{equation}
From the equalization between $\Gamma_d$ and $\omega_d$ at $k_d$, we find the expression of $k_d$ as 
(Eq. \eqref{eq: damrat} and \eqref{eq: gamlak})
\begin{equation}\label{eq: kdlak}
    k_d=\mathcal{C}^{-\frac{3}{4}} L^{-\frac{1}{4}} V_L^\frac{3}{4} \mathcal{E}_M^{-\frac{3}{4}}.
\end{equation}
We see that the damping scale increases with the growing magnetic energy $\mathcal{E}_M$.

The dynamo is in the linear regime, i.e., kinematic dynamo, as long as 
the magnetic energy is below the turbulent kinetic energy at $k_d$, 
\begin{equation}
     \mathcal{E}_M < \frac{1}{2} v_d^2.
\end{equation}
Combining the above relation with Eq. \eqref{eq: vkd} and \eqref{eq: kdlak} yields 
\begin{equation}\label{eq: pram2}
    \mathcal{E}_M < \frac{\mathcal{C}}{4}  L^{-1} V_L^3.
\end{equation}
It can be further rewritten as 
\begin{equation}\label{eq: rert}
    R_\text{ene}  <  \frac{\xi_n}{6} R_t,
\end{equation}
with 
\begin{equation}\label{eq: ratet}
   R_\text{ene} =  \frac{\mathcal{E}_M}{ \frac{1}{2} V_L^2} , ~~ R_t = \frac{\nu_{ni}^{-1}}{L/V_L}
\end{equation}
as the ratio between $\mathcal{E}_M$ and the turbulent energy at $L$ and 
the ratio between the neutral-ion collision time and the largest eddy-turnover time,
where the expression of $\mathcal{C}$ in Eq. \eqref{eq: damrat} is used.
When the magnetic energy satisfies both conditions in Eq. \eqref{eq: pram1} and \eqref{eq: pram2}, 
one should consider the damping regime for the kinematic dynamo growth of the magnetic field.

As illustrative examples, we use the typical ISM phases 
(see \citealt{Drai11}), e.g.,
the cold neutral medium (CNM) and molecular cloud (MC), as the preshock conditions for 
supernova remnants interacting with atomic and molecular clouds. 
Their typical parameters are listed in Table \ref{tab: cnm}
(see \citealt{Dra98}), 
including the number densities of the atomic hydrogen $n_H$ and electrons $n_e$, 
the temperature $T$.
For the turbulence parameters, 
we adopt the characteristic scale $\sim 0.1$ pc of the density structure in the CNM
\citep{Hei03}
and MC
\citep{go98,Mot07}
as the injection scale $L$ of the upstream turbulence. 
The turbulent velocity at $L$ depends on the density perturbation $\Delta \rho/ \rho$ and the shock speed $v_\text{sh}$
\citep{Dru12}. 
As an order-of-magnitude estimate, we adopt
\begin{equation}
     V_L \sim \frac{\Delta \rho}{\rho} v_\text{sh}.
\end{equation}
For an illustrative purpose,
we assume $\Delta \rho / \rho \sim 1$ and $V_L$ on the order of $10^3$ km/s in our following calculations.  
It is worthwhile to notice that the density perturbation can be vastly different in different ISM phases, 
e.g., local density enhancements with $\Delta \rho/ \rho \gg 1$ but a small volume filling factor in clumpy MCs 
(see \citealt{Stu88, HF12}),
the low density contrast with $\Delta \rho/ \rho \ll 1$ in more diffuse medium. 
Thus the corresponding $V_L$ can actually deviate significantly from $v_\text{sh}$.

We find that given the magnetic field strength $B_0$ in the ambient ISM, 
the initial magnetic energy is 
\begin{equation}
\mathcal{E}_{M0} = \frac{1}{2} V_{A0}^2 = \frac{B_0^2}{8\pi\rho},
\end{equation}
where $V_{A0}$ is the Alfv\'{e}n speed. 
It satisfies 
\begin{equation}
          \mathcal{C}^{-1} \nu_n < \mathcal{E}_{M0} < \frac{\mathcal{C}}{4}  L^{-1} V_L^3 ,
\end{equation}
or in terms of the ionization fraction $\xi_i = \rho_i/\rho$, 
\begin{equation} \label{eq: pram1ion}
   \xi_i < \frac{\xi_n \mathcal{E}_{M0}}{3 \gamma_d \rho \nu_n },  ~  \xi_i < \frac{\xi_n V_L^3}{12 \gamma_d \rho L \mathcal{E}_{M0}},
\end{equation}
for both the CNM and MC, 
indicative of the dominant ion-neutral collisional damping and 
the kinematic dynamo growth of magnetic energy 
according to the above analysis (Eq. \eqref{eq: pram1} and \eqref{eq: pram2}). 
In particular, we find $R_\text{ene} \approx 4\times10^{-6}$, $R_t \approx 185$
in the case of the CNM 
and $R_\text{ene} \approx 4\times10^{-7}$, $R_t \approx 6$
in the case of the MC (Eq. \eqref{eq: ratet}).
Naturally, $R_\text{ene}$ is very small as the turbulence in the precursor carries substantial kinetic energy originating from the supernova ejecta,
while a large $R_t$ suggests the weak coupling between neutrals and ions and thus strong damping. 
Here we also adopt 
$m_i=m_n=m_H$ for the masses of ions and neutrals in the CNM 
and $m_i = 29 m_H$, $m_n = 2.3 m_H$ for those in the MC
\citep{Shu92}, 
$\gamma_d = 3.5\times 10^{13}$ cm$^3$g$^{-1}$s$^{-1}$ 
\citep{Drai83},
and $\sigma_{nn} \approx 10^{-14}$ cm$^2$
\citep{VrKr13}.

\begin{table*}[t]
\centering
\begin{threeparttable}
\caption[]{Parameters in the preshock region}\label{tab: cnm} 
  \begin{tabular}{ccccccccc}
     \toprule
              &  $n_H [\text{cm}^{-3}]$ &  $n_e/n_H$  &   T [K]        & $B_0$ [$\mu$ G]     
              &  $k_\nu^{-1}$ [pc]   & $k_{d0}^{-1}$ [pc]       &  $t_\text{dyn}$ [yr] &  $B_\text{dyn}$ [$\mu$ G]  \\
      \hline
   CNM  &  $30$                           &  $10^{-3}$    &    $100$     &  $5$                           
             &   $1.3\times10^{-7}$ & $1.2 \times10^{-4}$  & $741.9$   & $452.6$      \\
   MC     &  $300$                         &  $10^{-4}$    &   $20$        & $5$     
             &  $1.7\times10^{-8}$  &  $1.6\times10^{-6}$  & $749.7$  &  $7.7\times10^3$ \\
    \bottomrule
    \end{tabular}
 \end{threeparttable}
\end{table*}

\subsection{Dynamo evolution of the magnetic field}

The kinematic dynamo is described by the Kazantsev theory, with 
$\mathcal{E}_M$ determined by the integral of the Kazantsev spectrum 
\citep{Kaza68, KulA92, Sch02, Bran05}
over the wavenumbers smaller than $k_d$, 
\begin{equation}\label{eq: enlak}
\begin{aligned}
  \mathcal{E}_M
  & = \frac{1}{2} \int_0 ^ {k_d} M(k,t) dk \\
  & = \frac{1}{2} \int_0 ^ {k_d} M_0 \exp{\bigg(\frac{3}{4} \int \Gamma_d dt \bigg)} \bigg( \frac{k}{k_0} \bigg)^\frac{3}{2} dk \\
                     & = \frac{1}{5} M_0 k_0  \exp{\bigg(\frac{3}{4} \int \Gamma_d dt \bigg)} \Big(\frac{k_d}{k_0}\Big)^\frac{5}{2},
\end{aligned}
\end{equation}
where $M(k,t)$ is the growing Kazantsev spectrum of magnetic energy,
and $M_0$ is the initial magnetic energy spectrum centered about $k_0$. 
By combining Eq. \eqref{eq: gamlak}, \eqref{eq: kdlak}, and \eqref{eq: enlak}, after some algebra we can reach the evolution law of the magnetic energy, 
\begin{equation}\label{eq: enetev3}
      \sqrt{\mathcal{E}_M} = \sqrt{\mathcal{E}_{M0}} + \frac{3}{23} \mathcal{C}^{-\frac{1}{2}} L^{-\frac{1}{2}} V_L^\frac{3}{2} t,
\end{equation}
where $\mathcal{E}_{M0}$ is the initial magnetic energy.
This dynamo regime has been 
identified as the damping kinematic dynamo in XL16.
Since $B = \sqrt{8 \pi \rho \mathcal{E}_M}$, it shows that the magnetic field strength grows linearly with time. 
The time evolution of $k_d$ can also be obtained by inserting the above equation into Eq. \eqref{eq: kdlak},
\begin{equation}\label{eq: kddarap}
    k_d = \Big[k_{d0}^{-\frac{2}{3}}+\frac{3}{23}L^{-\frac{1}{3}}V_L t\Big]^{-\frac{3}{2}},
\end{equation}
where $k_{d0}$ is the initial damping scale corresponding to $\mathcal{E}_{M0}$.

The kinematic saturation happens when the magnetic energy grows to equipartition with the local turbulent energy, i.e., 
\begin{equation}
    \mathcal{E}_M = \frac{1}{2} v_d^2. 
\end{equation}
By using Eq. \eqref{eq: vkd} and \eqref{eq: kdlak}, the above equation gives the 
critical damping scale $k_{d,\text{cr}}$ corresponding to the kinematic saturation,  
\begin{equation}\label{eq: cridams}
   k_{d,\text{cr}} = \Big(\frac{\mathcal{C}}{2}\Big)^{-\frac{3}{2}} L^\frac{1}{2} V_L^{-\frac{3}{2}}  .
\end{equation}
For the CNM and MC under consideration, there is 
\begin{equation}\label{eq: para3}
     k_{d,\text{cr}}  L = \Big(\frac{2L}{\mathcal{C}V_L}\Big)^\frac{3}{2} <1,
\end{equation}
or equivalently, 
\begin{equation}
      \xi_i < \frac{\xi_n V_L}{6  \gamma_d \rho L}. 
\end{equation}
It implies that the kinematic saturation cannot be reached during the entire dynamo process.

When the damping scale increases up to $L$,
we have the timescale of the damping 
kinematic dynamo in the partially ionized gas (Eq. \eqref{eq: kddarap})
\begin{equation}\label{eq: tdyn}
   t_\text{dyn} = \frac{23}{3} L^\frac{1}{3} V_L^{-1} (L^\frac{2}{3} - k_{d0}^{-\frac{2}{3}}),
\end{equation}
which is approximately $7.7$ largest eddy-turnover times. 
At $t_\text{dyn}$, the magnetic energy is amplified to (Eq. \eqref{eq: kdlak}, \eqref{eq: enetev3}, and \eqref{eq: tdyn})
\begin{equation}
    \mathcal{E}_\text{dyn} =  \mathcal{C}^{-1} L V_L.
\end{equation}
The calculated $t_\text{dyn}$ and the field strength corresponding to $\mathcal{E}_\text{dyn}$ in the CNM and MC
are presented in Table \ref{tab: cnm}.
The results here only serve as an order-of-magnitude estimate. 
More rigorous calculations rely on a more realistic and detailed modeling of the shock, 
as well as the ambient environment. 
We caution that with the development of the strong magnetic field in front of the shock, the shock 
Alfv\'{e}n Mach number $M_{\text{sh},A} = v_\text{sh}/V_A$ decreases. 
In the linear kinematic dynamo regime, the magnetic back-reaction on the shock propagation is usually unimportant. 
But in the case when the Alfv\'{e}n speed corresponding to the amplified magnetic filed at a later time of the dynamo
approaches the shock speed (e.g. in the case of MC),
the strong shock jump condition will not be satisfied. Consequently, 
the turbulence driving and the dynamo growth also cease.
In the self-regulated system in the realistic situation, 
the dynamo efficiency decreases with the arising of the dynamically important magnetic field.

Moreover,
for the amplification of the upstream magnetic field, 
the turbulent dynamo can only operate  
within the precursor crossing time $t_c = L_p/v_\text{sh}$. 
The maximum size of the CR precursor $L_p$ is determined by the diffusion of the highest-energy CRs, 
\begin{equation}
    L_p = \frac{\kappa_\text{max}}{v_\text{sh}}, 
\end{equation}
where the CR diffusion coefficient $\kappa$ should be computed self-consistently as they  
interact with the magnetic fluctuations amplified by the CR-driven turbulence.

\subsection{Magnetic field structure and CR diffusion}\label{ssec: msdf}

\begin{figure*}[htbp]
\centering
\subfigure[Preshock]{
   \includegraphics[width=8cm]{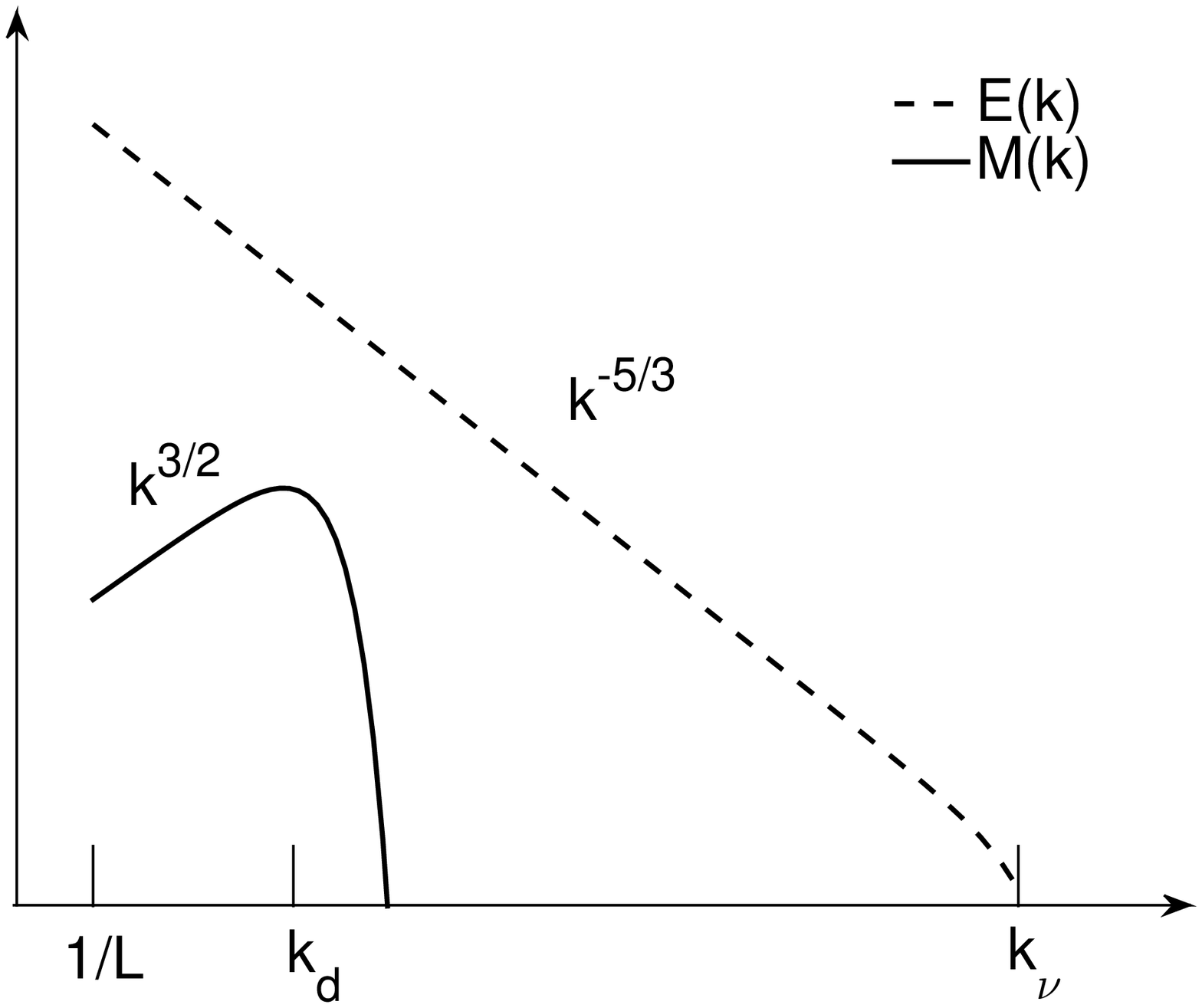}\label{fig: dams}}
\subfigure[Postshock]{
   \includegraphics[width=8cm]{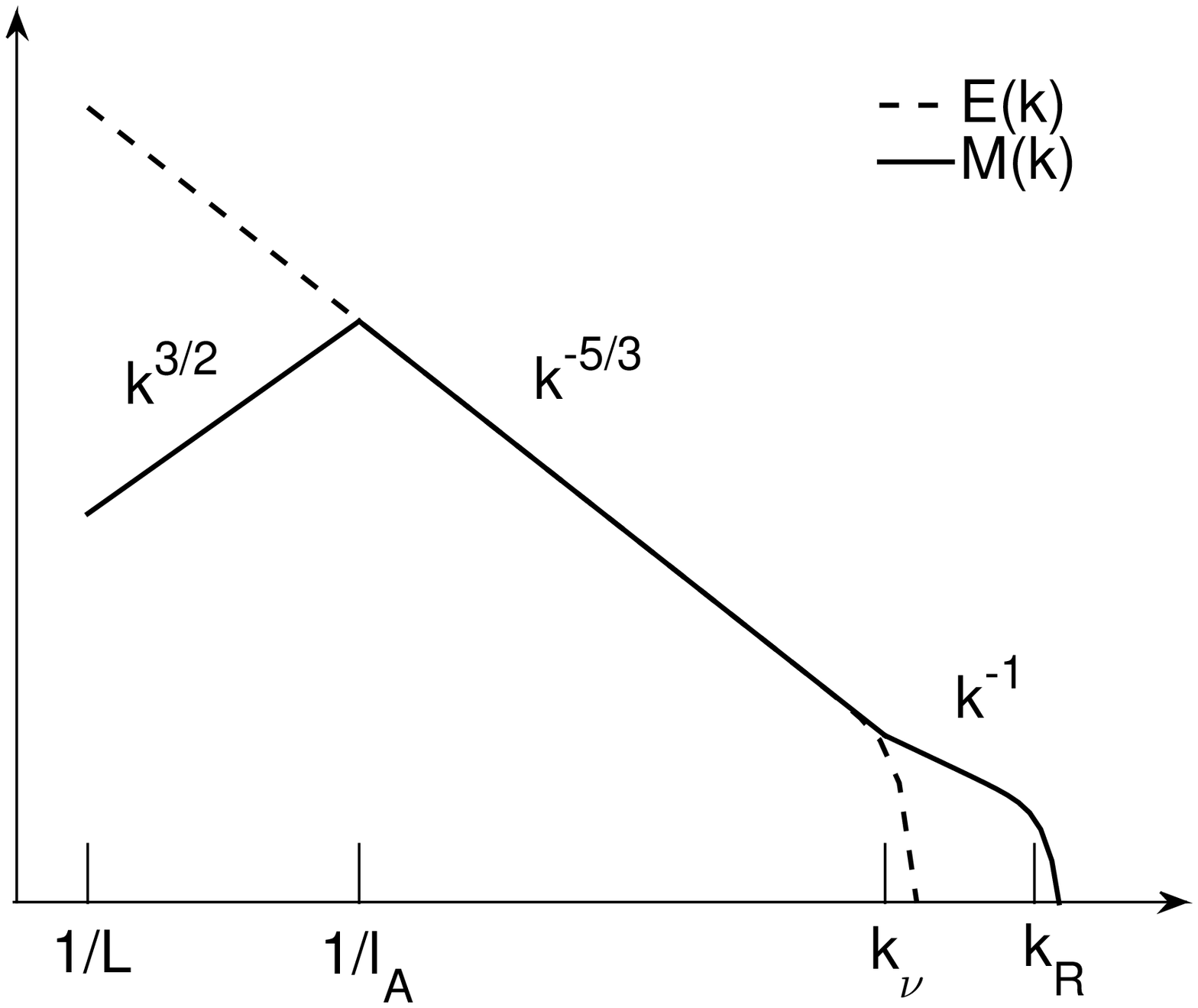}\label{fig: nons}} 
\caption{ Sketches of the magnetic (solid line) and turbulent (dashed line) energy spectra during the dynamo processes 
in the (a) preshock and (b) postshock regions.
Fig. \ref{fig: dams} is taken from \citet{XLr17}.}
\label{fig: sket}
\end{figure*}

Fig. \ref{fig: dams} illustrates the magnetic energy spectrum $M(k)$ during the damping kinematic dynamo. 
The Kazantsev spectrum extends over 
large scales down to the spectral peak at the damping scale, 
where the spectrum is cut off due to ion-neutral collisional damping. 
Since the magnetic energy is well below the turbulent kinetic energy over all scales, 
the nonlinear Lorentz back-reaction on turbulent motions is insignificant, 
and the entire dynamo process stays in the linear regime.

The magnetic field resulting from the damping kinematic dynamo in the presence of ion-neutral collisions 
has a characteristic scale as the increasing damping scale. 
It means that the magnetic field has a tangled structure over larger scales, 
but is smooth over smaller scales with the magnetic fluctuations suppressed. 
The bending of field lines by the 
super-Alfv\'{e}nic turbulent motions at $k_d$ gives rise to an effective mean free path $\lambda_\text{mfp}$ of CRs equal to $k_d^{-1}$
(see \citealt{Brunetti_Laz}).
Note that the Larmor radius $r_L$ of CRs should be smaller than $k_d^{-1}$. 
The corresponding diffusion coefficient is 
\begin{equation}
    \kappa \approx \frac{c \lambda_\text{mfp}}{3} = \frac{c}{3 k_d}, 
\end{equation}
where $k_d$ is given by Eq. \eqref{eq: kddarap}. 
Suppose that the damping scale, as well as $\lambda_\text{mfp}$, can reach $L$, thus the precursor has the maximum thickness 
\begin{equation}
    L_p = \frac{c L}{3v_\text{sh}}, 
\end{equation}
and the crossing time is 
\begin{equation}
    t_c  = \frac{c}{3 v_\text{sh}} \frac{L}{v_\text{sh}} = \frac{c}{3 v_\text{sh}} \frac{V_L}{v_\text{sh}} \frac{L}{V_L}. 
\end{equation}
If we adopt $V_L \sim v_\text{sh} \sim 10^3$ km/s, 
then $t_c$ is equal to $100$ largest eddy-turnover times, 
which is much larger than $t_\text{dyn}$ (Eq. \eqref{eq: tdyn}) and certainly 
sufficient for the full development of the precursor turbulence and the damping kinematic dynamo. 
More conservatively, the turbulent velocity should satisfy 
$V_L / v_\text{sh} > 7.7\times10^{-2}$ (see Eq. \eqref{eq: tdyn})
for the dynamo to proceed till the outer scale of the turbulence.

At the Bohm limit with $r_L = L$, the maximum energy of CRs whose effective diffusion is governed  
by the amplified magnetic field is 
\begin{equation}\label{eq: conce}
    E_\text{CR, max} = e B_\text{dyn} L ,
\end{equation}
where $e$ is the particle's charge.
Given the field strength in Table \ref{tab: cnm}, it is 
$4.2\times10^{16}$ eV in the CNM and 
$7.1\times10^{17}$ eV in the MC. 
It implies that the damping kinematic dynamo in the precursor can easily generate the magnetic field 
required for the acceleration of CRs up to the knee energy of $\sim 10^{15}$ eV and beyond.

The magnetic field structure and the related CR diffusion in the preshock region are critical for the DSA. 
The limits on the DSA arising from the partial ionization of the upstream medium 
were studied by 
e.g. \citealt{Dru96, Mal11}, 
under the consideration of the ion-neutral collisional damping of Alfv\'{e}n waves. 
In the situation with the precursor turbulence, 
the turbulent dynamo increases both the strength and length scale of the magnetic field
\footnote{Notice that with the magnetic field intensified through the turbulent dynamo, 
the cutoff scales of Alfv\'{e}n waves increase (see e.g. \citealt{Xuc16}).}.
Starting from the weak interstellar field strength and in the presence of the severe damping, 
the turbulent dynamo generates the damping-scale magnetic field. 
Without the pitch-angle scattering, 
CRs with the Larmor radii smaller than the damping scale gyrate about the field lines. 
The random change of the magnetic field orientation over the distance equal to the damping scale 
entails an effective diffusion of CRs. 
Unlike the resonant interaction, both low- and high-energy CRs undergo the same diffusion process
in this dynamo-generated magnetic field
\footnote{The low-energy CRs can also be subject to additional confinement via the scattering with 
e.g. the current-driven instability 
\citep{Bell2004}
on small scales. }.

The arising of the precursor turbulence depends on the density fluctuations in the upstream medium, 
which have been commonly observed in the ISM
\citep{Armstrong95,CheL10,XuZ17},
and especially in the cold and dense phases 
\citep{Laz09rev,HF12}
but with a small volume filling factor
\citep{Tie,HS13}. 
In the case that the SNR shock propagates through a relatively uniform ambient medium, 
the above dynamo process would become less efficient in the mildly turbulent precursor.

\section{Magnetic field amplification in the fully ionized postshock region}
\label{sec: post}

SNR shocks efficiently heat the ISM gas, resulting in a high temperature in the postshock region. 
We assume that 
the partially ionized gas passing through the shock becomes fully ionized downstream. 
Compared with the partially ionized preshock medium, the magnetic fluctuations in the fully ionized 
postshock medium are only marginally damped by the resistivity, 
with the resistive scale much smaller than the viscous scale (see the Appendix). 
On the other hand, 
the equipartition scale with the local turbulent energy in balance with the initial magnetic energy is given by 
\citep{Lazarian06}
\begin{equation}
\begin{aligned}
    l_A &= L M_A^{-3} = L V_L^{-3} V_A^3  \\ 
          &= 1.3\times10^{13} \Big(\frac{L}{0.1 \text{pc}}\Big) \Big(\frac{V_L}{100 \text{km/s}}\Big)^{-3}   \\
          &  ~~~~\Big(\frac{n_i}{10 \text{cm}^{-3}}\Big)^{-\frac{3}{2}}    \Big(\frac{B}{5 \mu \text{G}}\Big)^3 \text{cm} ,
\end{aligned}
\end{equation}
where $M_A$ is the Alfv\'{e}n Mach number, and the ion number density $n_i = n_H$. 
Because it is larger than the viscous scale of turbulence (Eq. \eqref{eq: knui}), 
the turbulent dynamo falls in the nonlinear regime 
in the presence of the significant Lorentz back-reaction on the turbulent motions over smaller scales.
The initial field strength adopted here is comparable to the interstellar value. 
When the precursor dynamo is also taken into account, a larger $l_A$ is expected, 
and the postshock dynamo starts with a stronger magnetic field.

The magnetic energy spectrum $M(k)$ during the nonlinear dynamo process is illustrated in Fig. \ref{fig: nons}. 
The Kazantsev spectrum exists on scales above $l_A$ where the turbulence is super-Alfv\'{e}nic.
The trans-Alfv\'{e}nic MHD turbulence is developed over smaller scales, with the same Kolmogorov form for both the 
turbulent and magnetic energy spectra. 
Along with the magnetic energy growth, the equipartition scale $l_A$ shifts to ever larger scales and eventually 
the MHD turbulence extends up to $L$. 
In the sub-viscous range, i.e., $k>k_\nu$, $M(k)$ is further prolonged to the resistive cutoff 
and follows the $k^{-1}$ profile as a result of the balance between the magnetic tension force and the viscous drag
(\citealt{CLV_newregime, CLV03, LVC04}; XL16). 
The numerical evidence for the above scalings of $M(k)$ can be found in dynamo simulations, 
e.g., \citet{Hau04, Bran05}.

Unlike the kinematic dynamo which is only subject to the microscopic magnetic diffusion, e.g., ambipolar diffusion in a partially ionized medium, 
the nonlinear dynamo mostly suffers the turbulent magnetic diffusion 
arising in the MHD turbulence. 
Since the turbulent diffusion rate is comparable to the dynamo growth rate, i.e., the eddy-turnover rate
\citep{LV99}, 
the nonlinear dynamo is inefficient. 
The evolution law of the magnetic energy in the nonlinear regime with the turbulent diffusion taken into account 
was analytically derived by XL16, 
\begin{equation}\label{eq: evolaw}
       \mathcal{E}_M = \mathcal{E}_{M0} + \frac{3}{38} \epsilon t, 
\end{equation}
where  
\begin{equation}
     \epsilon = L^{-1} V_L^3
\end{equation}
is the constant turbulent energy transfer rate. 
It reveals a linear-in-time growth of magnetic energy with the growth rate as a small fraction of $\epsilon$, 
which is consistent with direct numerical measurements, e.g., 
\citet{CVB09},
\citet{Bere11}.

When the dynamo saturation is achieved at $L$, the magnetic energy is equal to the kinetic energy of the largest turbulent eddy,
\begin{equation} \label{eq: finsat}
    \mathcal{E}_M = \frac{1}{2} V_L^2. 
\end{equation}
The timescale of the nonlinear dynamo is (Eq. \eqref{eq: evolaw})
\begin{equation}\label{eq: durnod}
   \tau_\text{nl}  =  \frac{38}{3 \epsilon}  ( \mathcal{E}_M - \mathcal{E}_{M0}).
\end{equation}

To examine the applicability of the above analysis on the magnetic field amplification in the postshock region, we next
carry out a comparison between our theoretical predictions and the numerical results in 
\citet{Ino09} (hereafter I09). 
They performed MHD simulations 
of SNR shocks propagating through the inhomogeneous ISM.
We only consider parallel shocks (Model 2, 3, and 4 in I09) and disregard 
additional magnetic field amplification due to the shock compression. 
The parameters that we use are listed in Table \ref{tab: posh}.

\begin{table}[t]
\centering
\begin{threeparttable}
\caption[]{Parameters of the postshock medium}\label{tab: posh} 
  \begin{tabular}{ccccc}
     \toprule
   &              $n_i$ [cm$^{-3}$]          &  $B_0$ [$\mu$ G]            &  $L$ [pc]              &  $V_L$ [km/s]     \\       
   \hline                                                          
   Model 2   &    $10$                           &  $6$                                 &  $0.1$                  &  $450$                \\
   Model 3   &    $19$                           &  $6$                                 &  $0.1$                  &  $240$                \\
   Model 4   &    $35$                           &  $6$                                 &  $0.1$                  &  $130$               \\
    \bottomrule
    \end{tabular}
 \end{threeparttable}
\end{table}

\begin{figure*}[htbp]
\centering
\subfigure[Temporal evolution of $B$]{
   \includegraphics[width=8cm]{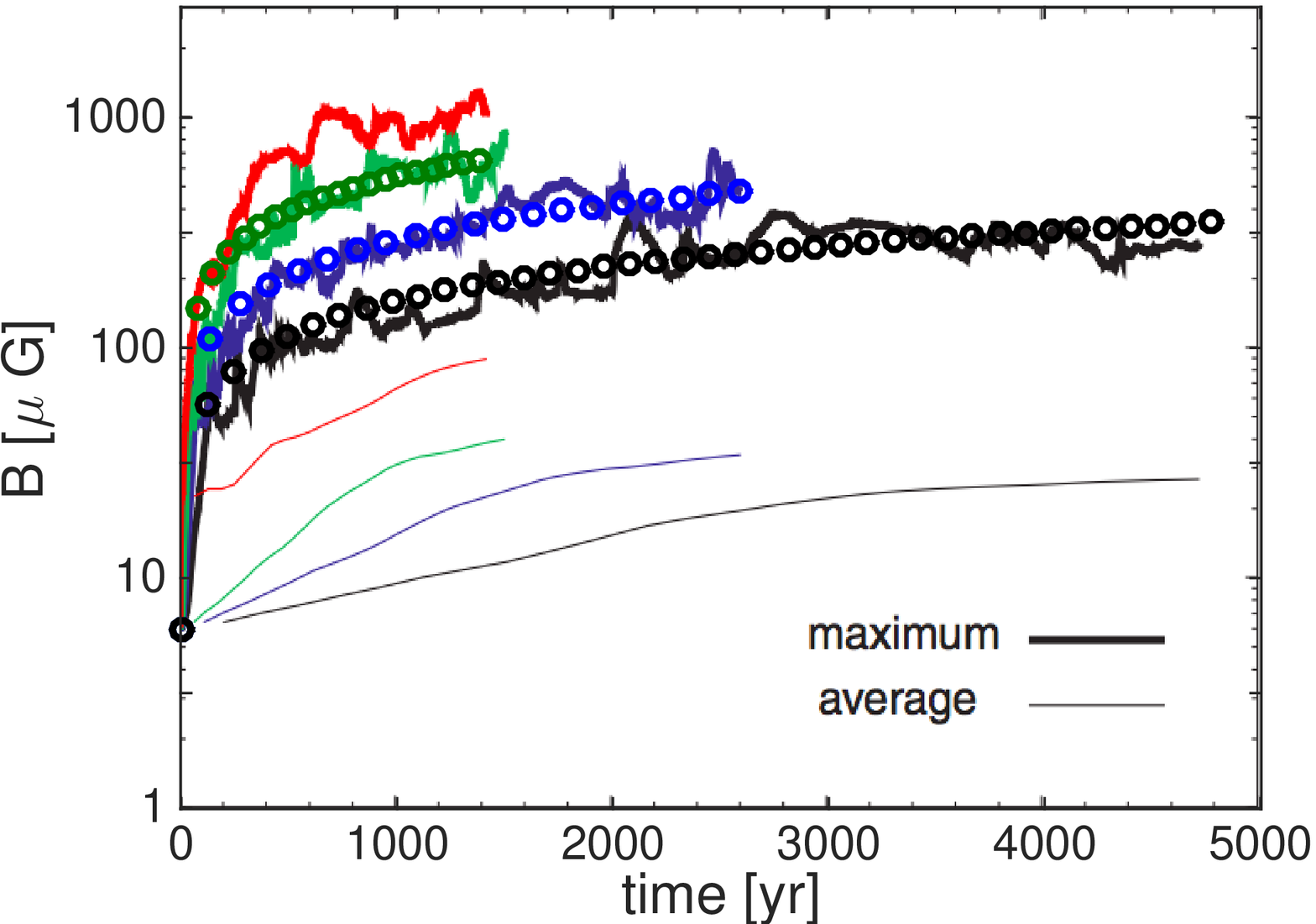}\label{fig: sh1}}
\subfigure[Spatial profiles of $B$]{
   \includegraphics[width=8cm]{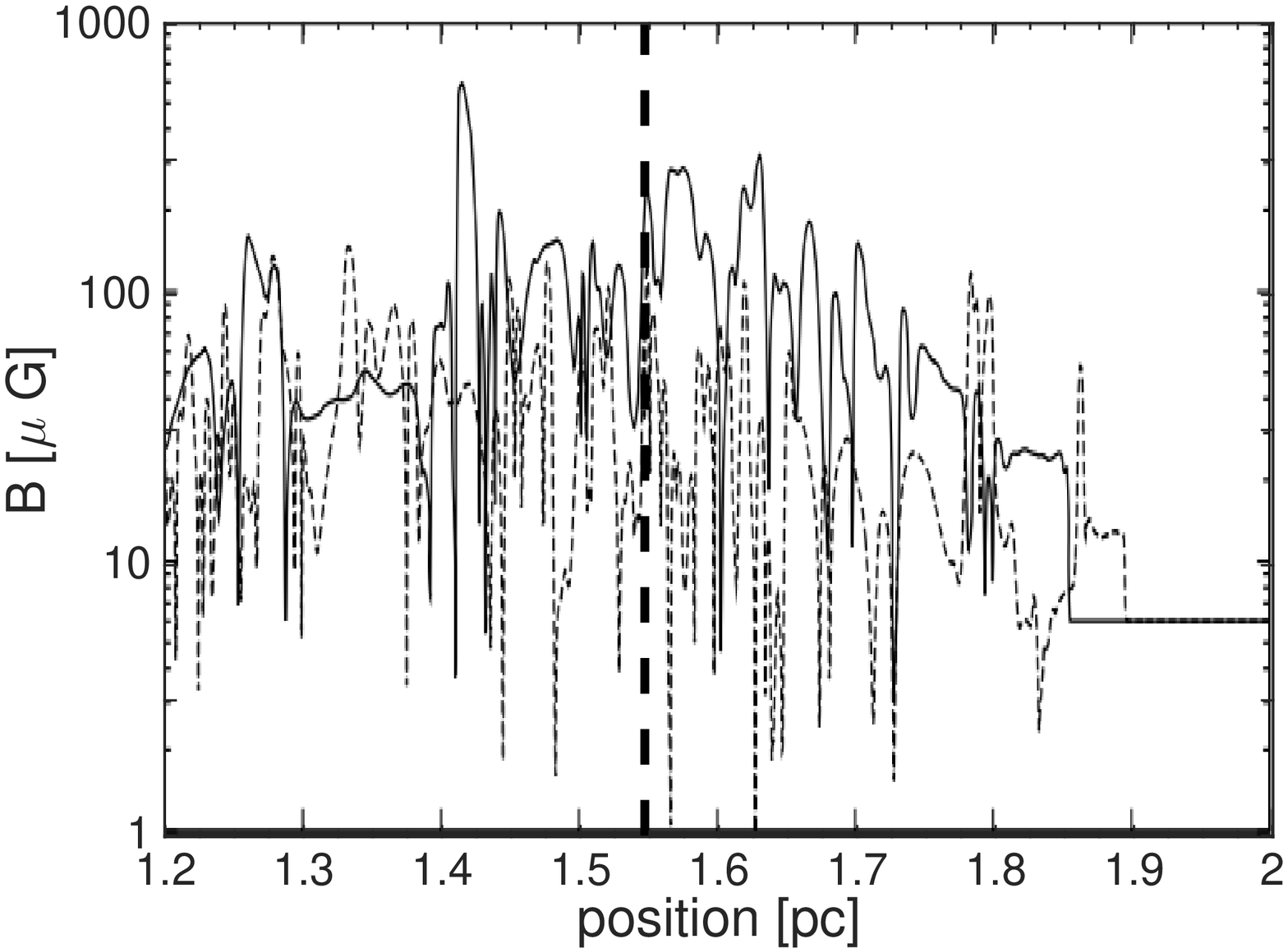}\label{fig: sh2}}  
\caption{ Comparisons between our analyses and the numerical results from I09. 
(a) Evolution of maximum (thick) and average (thin) field strengths. 
(b) Spatial profile of the field strength for perpendicular (solid) and parallel (dotted) shocks. 
On the original figures taken from I09,
the overplotted circles in (a) represent our analytical field strength in comparison with different numerical models 
(Models 2 (green), 3 (blue), and 4 (black))
for a parallel shock.  
The overplotted vertical dashed line in (b) shows our estimated position for the saturation of the amplified magnetic field in the postshock region of 
a parallel shock (Model 2, dotted line).
}
\label{fig: dynsh}
\end{figure*}

Starting from the initial field strength $B_0$, the temporal evolution of the magnetic field is calculated according to Eq. \eqref{eq: evolaw}. 
As shown in Fig. \ref{fig: sh1}, our analytical results can well match the numerical findings.
The saturated field strength is determined by the turbulent velocity $V_L$ (Eq. \eqref{eq: finsat}). It approaches $1$ mG given the settings of 
Model 2.

We can also calculate the distance between the location where the magnetic field reaches saturation and the shock front, which is given by 
\begin{equation}\label{eq: dissat}
     l_\text{sat} \approx v_\text{down} \tau_{nl}, 
\end{equation}
where the downstream bulk velocity is 
$v_\text{down} = 1/4 v_\text{sh}$ for a strong shock, and $v_\text{sh} = 1289$~km s$^{-1}$ is the shock velocity (Model 2 in I09). 
The timescale of the nonlinear dynamo $\tau_\text{nl}$ is provided by Eq. \eqref{eq: durnod}. 
Fig. \ref{fig: sh2} shows that 
our estimate (the vertical dashed line) coincides with the peak position of the magnetic field profile numerically produced in I09.

Because of the inefficiency of the nonlinear turbulent dynamo and its prolonged timescale, the peak field strength is developed at a 
significant distance, rather than the immediate vicinity, behind the shock front. 
This feature naturally explains the location of the X-ray hot spots detected at more than $0.1$ pc downstream of SNR shocks 
(I09;  \citealt{Uch07, Uch08}).

The above comparison confirms the predictive power of the XL16 model for 
the nonlinear turbulent dynamo. 
It provides an analytically tractable means of studying the evolution and distribution of the magnetic field in the postshock region. 
Notice that in the case of efficient precursor dynamo, 
starting from a relatively high magnetization, 
the timescale of the postshock nonlinear dynamo can be much shortened, 
and the growth of the magnetic energy is insignificant.

The amplified magnetic field affects the diffusion of CRs behind the shock. 
For lower-energy CRs with 
$E_\text{CR} < e B(l_A) l_A$, 
where $B(l_A)$ is the field strength reached at the energy equipartition, 
their mean free path is independent of energy and given by $l_A$
(see BJL09).
The corresponding diffusion coefficient is $\kappa = c l_A/3$. 
For higher-energy CRs with the Larmor radius exceeding $l_A$, 
the larger-scale magnetic fluctuations following the Kazantsev spectrum, i.e., $B(k) \propto k^{5/4}$,
are too weak to confine the CRs
as $eB(k)k^{-1} \propto k^{1/4}$.
But since $l_A$ increases with time, more energetic CR particles get influenced 
by the amplified magnetic field 
during the nonlinear turbulent dynamo.
At the distance $l_\text{sat}$ (Eq. \eqref{eq: dissat}) 
behind the shock front, $l_A$ reaches the outer scale of turbulence $L$ at the nonlinear saturation, 
where CRs with $E_\text{CR} < e B(L) L$ have $\kappa = c L /3$.

As regards the CR diffusion in the direction perpendicular to the mean magnetic field, 
lower-energy CRs with 
$E_\text{CR} < e B(l_A) l_A$ 
are characterized by the fast superdiffusion, i.e. $\langle\sigma_\perp^2\rangle \propto t^\alpha$, $\alpha>1$, 
where $\langle\sigma_\perp^2\rangle$ is the mean squared displacement in the perpendicular direction
\citep{LY14}.
The superdiffusion dominates the perpendicular transport of CRs as $l_A$ increases with the distance behind the shock front. 
The higher-energy CRs undergo the normal diffusion with 
$\langle\sigma_\perp^2\rangle$ as a linear function of time.

\section{Discussions}

Compared with earlier studies on the turbulent dynamo in the shock precursor,
e.g., BJL09, \citet{Del16},
based on the analytical results of turbulent dynamo in a partially ionized medium 
(XL16), 
we have demonstrated that the 
magnetic field amplification in a partially ionized medium can be drastically different from that in a fully ionized medium. 
In the CNM and MC, because of the severe damping, the dynamo remains in the linear regime with a linear-in-time growth of field strength, 
rather than an exponential growth expected for the kinematic dynamo in a highly ionized medium
\citep{Bal05},
or a linear-in-time growth of magnetic energy in the nonlinear regime 
(BJL09).
Through comparison with the current-driven instability
\citep{Bell2004}, 
BJL09 suggested that the nonlinear turbulent dynamo is more favorable in amplifying the upstream magnetic fields.
We present a more efficient dynamo regime  
arising in the partially ionized preshock medium than the nonlinear dynamo. 
Furthermore, different from the limited timescale considered in BJL09, 
the extended CR precursor in our scenario leads to an ample time for development of turbulence and magnetic field amplification
\footnote{We note that in our calculation for the maximum size of the precursor, we use the diffusion length scale of the 
highest-energy CRs. When confronting with the precursor thickness indicated by X-ray observations
(see \citealt{Vin12}),
one should adopt that corresponding to the electrons with 
the characteristic energy of their energy spectrum for the observable emission.}.
Our results support the notion that 
the turbulent dynamo in the shock precursor is the dominant process of magnetic field amplification.

It is important to stress that 
only when the ionization fraction is sufficiently high can the kinematic dynamo be considered unimportant compared with the nonlinear dynamo stage. 
In the new damping regime identified in XL16, depending on the ionization fraction, 
the kinematic dynamo can be important through the turbulence inertial range 
up to the equipartition scale, or the outer scale as the case discussed in the paper. 
Hence in a weakly ionized medium, the kinematic dynamo can have astrophysically important applications. 
In the preshock turbulence, 
the characteristic scale of the magnetic field grows with the increasing damping scale of turbulence, 
independent of the CR Larmor radius. 
The resulting damping-scale magnetic field regulates the   
diffusion behavior and thus the acceleration of CRs from low energies up to very high energies.
Different from the short-wavelength Bell mechanism, there is no need to invoke additional inverse cascade process 
for transporting the magnetic energy to larger scales.

In the turbulent postshock medium,
the observed year-scale X-ray variability of the compact hot spots 
suggests a largely amplified magnetic field of $1$ mG 
and very efficient particle acceleration in the emitting regions
\citep{Uch07, Uch08}. 
Our results on the amplification of the postshock magnetic field 
show that the analytical expectations in XL16 agree well with the existing numerical simulations. 
Depending on the postshock turbulent velocity, the saturated field strength of the nonlinear dynamo 
can be on the order of $1$ mG (Section \ref{sec: post}). 
As pointed out by I09, the electrons responsible for the hot spots are unlikely to be accelerated at the shock front in view of the 
fast synchrotron cooling.
The efficient in-situ acceleration is attributable to the reverse shock 
\citep{Uch08}
or secondary shocks arising from collisions of turbulent flows
(I09).
An alternative explanation could be the adiabatic non-resonant 
acceleration in MHD turbulence proposed by 
\citet{BruL16}
(see also \citealt{XZg17}).
In MHD turbulence, 
magnetic field lines can be stretched due to the turbulent dynamo and shrink via the turbulent reconnection, 
which happen through every eddy turnover
\citep{LV99}. 
The CRs with the Larmor radii much smaller than the size of the turbulent eddy
are attached to field lines and confined within the eddy during its eddy-turnover time, 
bouncing back and forth between converging magnetic fields in the turbulent reconnection region. 
This process is similar to the first-order Fermi acceleration
\citep{DeG05}. 
The resulting efficient acceleration within the eddy-turnover time may accommodate 
the year-scale variability in the synchrotron emission.
A detailed study on this mechanism and its applicability in CR acceleration in the postshock turbulence is the subject of an upcoming paper.

Our results on the magnetic field amplification and its implications on the diffusion of energetic particles in both the pre- and post-shock regions 
are important for further modelling the emission characteristics in comparison with multi-wavelength observations of SNRs 
(see e.g., \citealt{Zen17}).

\section{Summary}

By applying the turbulent dynamo theory formulated by XL16, 
we have investigated the magnetic field amplification in SNRs.

The dynamo evolution of magnetic fields in the pre- and post-shock media are very different. 
The dynamo in the weakly ionized upstream medium, e.g., the CNM and MC, 
is characterized by a linear-in-time growth of field strength, that is, 
the magnetic energy grows quadratically with time. 
It is slower than the exponential growth in the linear regime without damping, 
but faster than the linear-in-time growth of magnetic energy in the nonlinear regime.
In the extended CR precursor, 
the large damping-scale magnetic field formed at later times of the damping kinematic dynamo 
is beneficial for confinement of high-energy CRs. 
This finding is important for a 
realistic treatment of the shock acceleration of CRs in the partially ionized ISM.

Importantly, we provide the criterion for the dominance of 
the ion-neutral collisional damping over the neutral viscous damping 
and thus the emergence of the damping regime of the kinematic dynamo 
(Eq. \eqref{eq: pram1}).
Then the conditions for the entire dynamo process to be in the damping regime are given by 
Eq. \eqref{eq: pram2} (or Eq. \eqref{eq: rert}) and Eq. \eqref{eq: para3}.

The turbulent dynamo in the postshock region is in the nonlinear regime and drives a
linear-in-time and inefficient growth of magnetic energy. 
Provided a weak initial field strength in the postshock region,
the peak field strength at the dynamo saturation can only be reached in the farther downstream region. 
This explains the X-ray hot spots located far from the shock front. 
The consistency with the results of numerical simulations (e.g., I09)
shows that the XL16 analytical model for the nonlinear turbulent dynamo can be used to quantify the evolution and distribution 
of downstream magnetic fields.

The postshock magnetic turbulence can serve as an alternative source
besides the shock front 
for efficient acceleration of CRs. 
The corresponding acceleration mechanism deserves more attention and detailed analysis in future 
by confronting with updated observations. 
\\
\\

We thank the anonymous referee for insightful comments.
S.X. acknowledges the support for Program number HST-HF2-51400.001-A provided by NASA through a grant from the Space Telescope Science Institute, which is operated by the Association of Universities for Research in Astronomy, Incorporated, under NASA contract NAS5-26555.
A.L. acknowledges the support from grant
NSF DMS 1622353.

\appendix 
\label{app}

\section{The viscous and resistive scales in the fully ionized postshock region}

\begin{comment}
\begin{table*}[t]
\centering
\begin{threeparttable}
\caption[]{Parameters in the postshock region}\label{tab: posh} 
  \begin{tabular}{ccccccccccc}
     \toprule
     $n_i [\text{cm}^{-3}]$ &     T [K]        & $B_0$ [$\mu$ G]   & $L$ [pc] &  $V_L$ [km/s]  &  $l_A$ [pc]                 &  $\nu_{i,\perp}$ [cm$^2$ s$^{-1}$] & $\eta$ [cm$^2$ s$^{-1}$] &  $k_\nu^{-1}$ [cm]   & $k_R^{-1}$ [cm]       &  $P_m$  \\
      $10$                         &     $10^7$    &  $5$                       &  $0.1$    &  $10^2$           & $4.1\times10^{-6}$     & $2.0\times10^7$                        & $3.0\times10^2$              &    $4.0 \times10^{4}$     & $1.6 \times10^{2}$   & $6.7 \times10^4$     \\
    \bottomrule
    \end{tabular}
 \end{threeparttable}
\end{table*}
\end{comment}

In the presence of magnetic field, the ion viscosity is anisotropic. 
Even for the initial magnetization as low as the level in the ISM, the ion viscosity perpendicular to the field is very small
\citep{Sim55},
\begin{equation}
    \nu_{i,\perp}  = \frac{3k_BT \nu_{ii}}{10 \Omega_i^2 m_i}  
    = 2.0\times10^7  \Big(\frac{\ln \Lambda}{10}\Big)  \Big(\frac{T}{10^7 \text{K}}\Big)^{-\frac{1}{2}}  
       \Big(\frac{n_i}{10 \text{cm}^{-3}}\Big) \Big(\frac{B}{5 \mu \text{G}}\Big)^{-2} \text{cm}^2 \text{s}^{-1},
\end{equation}
where $c$, $m_e$, $e$, $\ln \Lambda$, $k_B$, $\nu_{ii}$, $\Omega_i$, $m_i (=m_H)$, $n_i$ are 
the speed of light, the electron mass and charge, the Coulomb logarithm, the Boltzmann constant, the ion collision and 
cyclotron frequencies, the ion mass (equal to the hydrogen atomic mass) and number density. 
It corresponds to a small viscous scale  
\begin{equation}\label{eq: knui}
    k_\nu = L^{-\frac{1}{4}} V_L^\frac{3}{4} \nu_{i,\perp}^{-\frac{3}{4}} 
              = 2.5\times10^{-5}  \Big(\frac{L}{0.1 \text{pc}}\Big)^{-\frac{1}{4}}  \Big(\frac{V_L}{100 \text{km/s}}\Big)^\frac{3}{4} 
               \Big(\frac{\nu_{i,\perp}}{2.0\times10^7 \text{cm}^2 \text{s}^{-1}}\Big)^{-\frac{3}{4}} \text{cm}^{-1} .
\end{equation}
On the other hand, the resistivity is 
\citep{Spit56},
\begin{equation}\label{eq: spiresis}
 \eta = \frac{c^2 \sqrt{m_e} e^2 \ln \Lambda}{4 (k_B T)^\frac{3}{2}}  
 = 3.0\times10^2 \Big(\frac{\ln\Lambda}{10}\Big) \Big(\frac{T}{10^7 \text{K}}\Big)^{-\frac{3}{2}} \text{cm}^2 \text{s}^{-1} .
\end{equation}
Then we have the magnetic Prandtl number,
\begin{equation}
   P_m = \frac{\nu_{i,\perp}}{\eta} = 6.7\times10^4 \Big(\frac{T}{10^7 \text{K}}\Big) \Big(\frac{n_i}{10 \text{cm}^{-3}}\Big) \Big(\frac{B}{5 \mu \text{G}}\Big)^{-2} ,
\end{equation}
exceeding unity by orders of magnitude. 
The resistive scale is given by 
\begin{equation}
     k_R = P_m^\frac{1}{2} k_\nu , 
\end{equation}
which is further shorter than the viscous scale.

\bibliographystyle{apj.bst}
\bibliography{yan}

\end{document}